\documentclass{JINST}
\let\ifpdf\relax

\usepackage[utf8]{inputenc}
\usepackage[T1]{fontenc}
\usepackage{textcomp}
\usepackage{nicefrac}
\newcommand{\um}[1][]{#1\,\textmu m}
\newcommand{\us}[1][]{#1\,\textmu s}
\newcommand{\keV}[1][]{\ensuremath{#1\,\mathrm{ke}\!\mathrm{V}}}

\def\Ed{\ensuremath{E_{\mathrm{d}}}}
\def\Et{\ensuremath{E_{\mathrm{t}}}}

\title{X-ray imaging with GEMs using 100\,\textmu m thick foils}
\author{H. Natal da Luz$^a$\thanks{Corresponding author.},~
J.A. Mir$^b$, X. Carvalho$^a$, J.M.F. dos Santos$^a$\\
\llap{$^a$}Atomic and Nuclear Instrumentation Group, Physics Department, University of Coimbra\\
  Rua Larga, PT3004-516 Coimbra, Portugal\\
\llap{$^b$}STFC Rutherford Appleton Laboratory,\\
  Chilton, Didcot, Oxon, OX11 0QX, UK\\
E-mail: \email{hugo@gian.fis.uc.pt}}

\abstract{A simple X-ray imaging system using off-the-shelf electronics and simple reconstruction algorithms aiming a spatial resolution of 1.7\,mm ($\sim 3\,\%$ of the detector length) is described in this work. For this, two 100\,cm$^2$ Gas Electron Multiplier (GEM) foils with a thickness of 100\,\textmu m (2-fold thicker than the standard ones) were immersed in a mixture of argon and carbon dioxide (70:30). The charge readout with 2D position determination was done with resistive charge division.

Due to their higher thickness with respect to the standard GEMs, the 100\,\textmu m thick GEM foils were found to be less prone to damage caused by the electrical discharges.

X-ray images are shown and some descriptions of the physical processes involved are presented. We describe the advantages of this method that allows counting each X-ray photon or particle entering the detector, its interaction position, as well as measuring of its energy. The results of our present work show a position resolution below 2\,mm, being limited by the gas mixture used, and not the detecting system, with a very good cost effectiveness. Future work is being carried out to optimize the present system for a medical application as a proton beam monitor.
}


\begin{document}

\section{Introduction}\label{sec:intro}
The Gas Electron Multiplier (GEM) is a Micropattern Gaseous Detector (MPGD) that has been successfully applied in Particle Physics and Medical Science, among other fields, for over a decade~\cite{Sau97}. A standard GEM configuration consists of \um[50] Kapton thick foil which is copper clad on both sides and perforated with \um[70] diameter holes in a \um[140] hexagonal pitch. By applying suitable electrical fields across the holes with the foil immersed in suitable a gas mixture, it is possible to multiply electrons from a primary ionization cloud and obtain a charge signal proportional to the energy deposited in the detector. The GEM can be cascaded with a second or a third multiplication stage until a satisfactory signal-to-noise ratio (SNR) is achieved. The two dimensional geometry of this kind of structure offers a very good solution whenever there are demands of large detection areas, such as the cases of position sensitive detectors for X-rays, particle beam monitors and others. In fact, the imaging capabilities of MPGDs have been well documented for GEMs~\cite{Bac02}, Micro-Hole and Strip Plates~\cite{Luz08} and other micropattern elements.

Most of the imaging applications with MPGDs have made use of discrete channel readout. This approach provides very good spatial resolutions, of the order of hundreds of \textmu m for areas as large as $100\,\mathrm{cm}^2$. However, it involves the use of a very large number of channels, and increases the complexity of the electronic system. Whenever a spatial resolution of the order of mm is needed, one can simplify the electronic readout and use resistive charge division, determining the position of interaction through the algorithms of the center of mass. However, this solution is very much dependent on obtaining a good signal-to-noise ratio. This means that the GEMs must operate at the highest gains possible, eventually too close to the discharge limit. The inevitable consequence of operating at such a regime is the higher probability of discharges resulting in permanent detector damage. The non-standard 100\,\textmu m thick GEM-foils, which have a two-fold thicker kapton\texttrademark{} suggest that a higher discharge power is needed to cause a short circuit accross the holes. This attribute is clearly advantageous towards the construction of a robust detector.

\section{Experimental Setup}\label{sec:exp}

\begin{figure}[tbp]
\centering
\includegraphics[width=.5\textwidth]{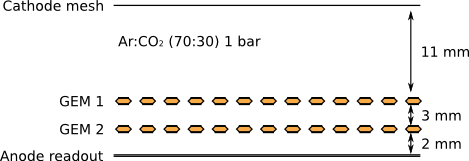}
\caption{Scheme of the detector, with the drift, transfer and induction gaps. The gas mixture was Ar:CO$_2$ (70:30) at 1 bar.}
\label{fig:esquema}
\end{figure}

GEM foils of $10\times 10\,\mathrm{cm}^2$ and a thickness of 100\,\textmu m were used in this work, in a double cascade, immersed in a mixture of argon:CO$_2$ (70:30) at atmospheric pressure. The drift, transfer and induction gaps were 11\,mm, 2.8\,mm and 2\,mm, respectively, as shown in figure \ref{fig:esquema}. The entrance window was in alluminized Mylar\texttrademark{} with a thickness of \um[25].

The charge readout electrode, with an area of $50\times 50\,\mathrm{mm}^2$, was composed of two layers of parallel strips with \um[200] pitch. The two layers were rotated by $90^\circ$ with respect to each other. The strips of each layer were interconnected by a resistive line and the signals were collected from both ends. The resistive line has the effect of dividing the charge to be collected at both ends. The ratio between the amount of charge reaching each side has information of the position of interaction for each dimension. This approach provides information of the position and and energy of each event produced in detector, making use of only four shaping/amplification channels.

In each of these four channels, the charge pulses were integrated by a charge sensitive preamplifier with a rise time of \us[10], and digitized by a CAEN VME1728 digital pulse processor, where the shaping and amplification was done. The amplitude and the time stamp, with a resolution of 10\,ns, of each pulse  were recorded in all the four electronic channels. In a first step of data processing to reconstruct the image, the pulses were arranged in groups of four, one in each channel, occurring within a time window of a few ns. Whenever this condition was not met, the pulses were discarded. The output of this first step was a collection of $(x,y,E)$ events, which composed the reconstructed image.

A copper X-ray tube was placed at a distance of about 2\,m from the detector window and several masks made with stainless steel or lead were used to project the images in the detector. The X-ray tube was filtered with a 1.5\,mm stainless steel sheet and the X-rays in the range 10--25\,\keV{} were used in the image reconstruction.

\begin{figure}[tbp]
\centering
\includegraphics[width=.4\textwidth]{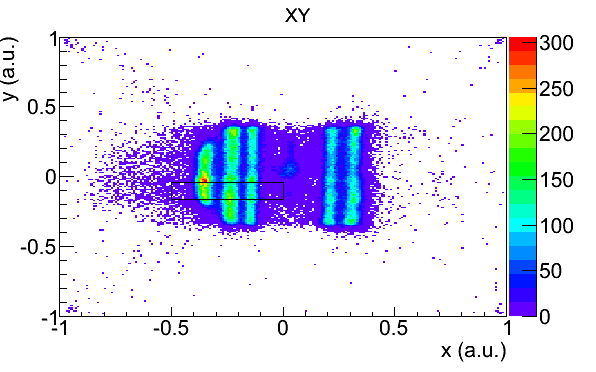}
\includegraphics[width=.4\textwidth]{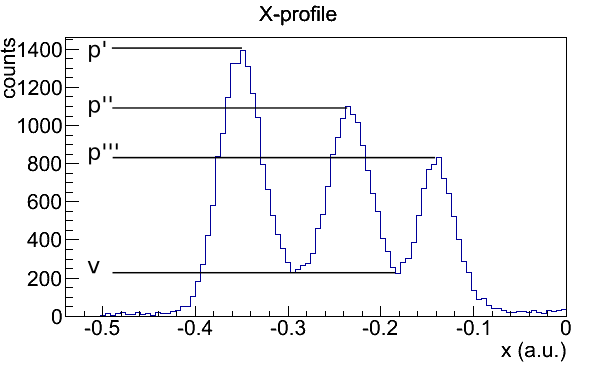}
\caption{Example of the determination of the contrast. The image on the right is the profile of the region marked in the image on the left. The average height of the peaks and the valleys is used to calculate the contrast of the image.}
\label{fig:cont}
\end{figure}

The drift and transfer fields were optimized by measuring the contrast of an array of slits with known pitch. To measure the contrast, the profile of an image of the slits was made and equation \ref{eq:cont} was used:

\begin{equation}\label{eq:cont}
C = \frac{p-v}{p},
\end{equation}

\noindent where $p$ is the average height of the three peaks and $v$ was the average height of the two valleys. The example in figure \ref{fig:cont} illustrates the procedure used to determine the contrast. This simple method does not correct for non-uniformities of the GEM foils and of the space between them, which causes fluctuations in the transfer and induction fields. The contrast is determined for a specific region of the detector. 

\begin{figure}[tbp]
\centering
\includegraphics[width=.4\textwidth]{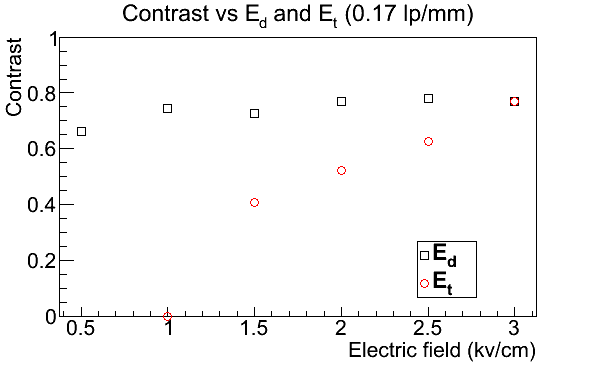}
\includegraphics[width=.4\textwidth]{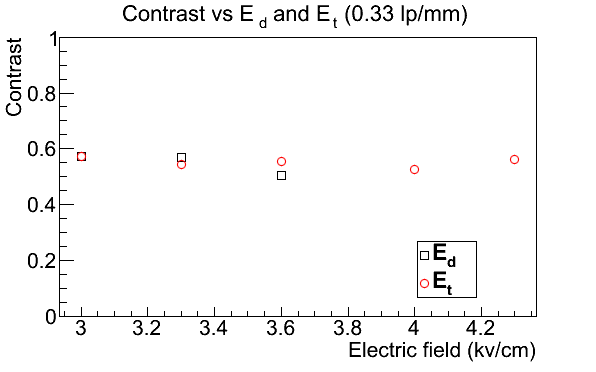}
\caption{Variation of the image contrast as a function of the drift and transfer electric fields (\Ed{} and \Et, respectively). Left: 0.17\,lp/mm slits. Right: 0.33\,lp/mm slits.}
\label{fig:fields}
\end{figure}

Figure \ref{fig:fields} shows the behavior of the contrast for different drift and transfer electric fields (\Ed{} and \Et, respectively). To obtain this data, whenever one of the fields were scanned, the other was kept at 3\,kV/cm (note the supperimposed data points at this value). For fields under 3\,kV/cm, slits of 0.17\,lp/mm (\emph{line pairs per mm}) were used and for the higher values, 0.33\,lp/mm were used. The drift field becomes optimal already at low values, as expected. It is also expected that, when the drift field is too high, some field lines intersect the copper surface of the GEM, between the holes. Consequently, some electrons from the primary do not cross the holes and are lost to the surface of the copper layer. The point at 3.6\,kV/cm suggests that this limmit is about to be reached.

In the transfer gap, the field must be above 3\,kV/cm to achieve the best contrast. This means that for lower fields, some electrons emerging from the first GEM are lost to the bottom electrode and do not reach the second GEM, resulting in a loss of signal and therefore decreasing the signal-to-noise ratio.
The drift, transfer and induction fields used throughout this work were 3, 3 and 4\,kV/cm, respectively.

\section{Results and discussion}\label{sec:res}

\begin{figure}[tbp]
\centering
\includegraphics[width=.6\textwidth]{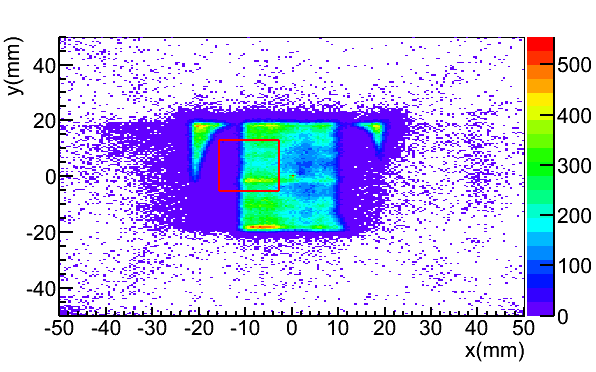}
\caption{The image of a stainless steel caliper with an aperture of 20\,mm. The region marked can be used to determine the Edge Spread Function.}
\label{fig:calip}
\end{figure}

To evaluate the \emph{point spread function} (PSF) of the system at one dimension, a stainless steel mask with a 1\,mm wide slit was imaged. This one dimensional approach of the PSF -- the \emph{line spread function} (LSF) -- is very convenient, because it is easier to image a slit than a pin-hole. The \nicefrac{pixels}{mm} value of the system was determined by imaging a caliper with an aperture of 20\,mm (figure \ref{fig:calip}) and measuring the number of pixels between the two edges of the profile.

\begin{figure}[tbp]
\centering
\includegraphics[width=.6\textwidth]{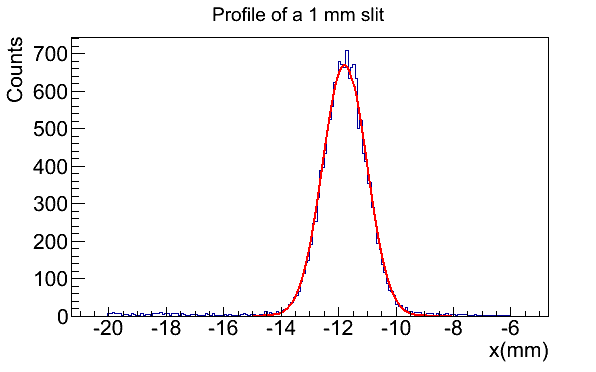}
\caption{Image of a 1\,mm slit. Its width is 1.78\,mm and gives an idea of the position resolution of the system.}
\label{fig:slit}
\end{figure}

The profile obtained for the 1\,mm slit is shown in figure \ref{fig:slit}. The width of the distribution is 1.78\,mm. This image is the convolution of the LSF with a rectangle with a width of 1\,mm, which is the slit. The contribution of the width of the slit to this profile tends to vanish as it becomes narrower. The fact that it appears wider in the image shows that the resolution of the system is above 1\,mm. To deconvolute the contribution of the slit and the LSF in this image equation \ref{eq:dec} is used \cite{Kle82}:

\begin{equation}
  \label{eq:dec}
  \sigma_x = w_{\mathrm{o}}\times\sqrt[3]{\left(\frac{w_{\mathrm{i}}}{w_{\mathrm{o}}}\right)^3-1},
\end{equation}

\noindent where $\sigma_x$ is the width of the LSF, $w_{\mathrm{o}}$ is the width of the slit and $w_{\mathrm{i}}$ is the width of the image. It is shown that the minimum width possible to image with this system is then 1.67\,mm.

\begin{figure}[tbp]
\centering
\includegraphics[width=.6\textwidth]{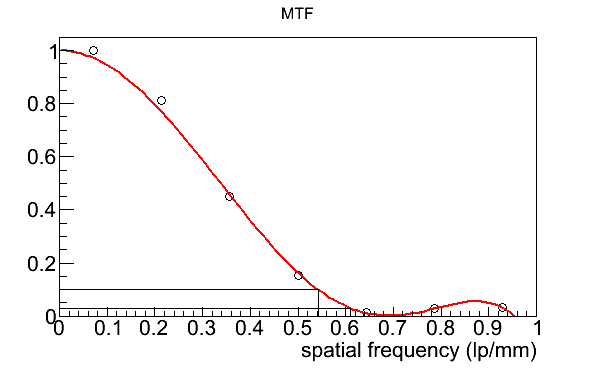}
\caption{The Modulation Transfer Function of one of the edges of figure \protect\ref{fig:calip}}
\label{fig:MTF}
\end{figure}

Figure \ref{fig:calip}, can also be used to further evaluate the imaging capabilities of this detector. The \emph{edge spread function} (ESF) of the region marked with a red rectangle has enough information to determine the behavior of the system. By deriving the ESF, one obtains the LSF and its \emph{discrete Fourier transform} is the \emph{modulation transfer function} (MTF). The MTF shows the behavior of the imaging system as a function of the imaged objects. Its value is the contrast for line pairs with the corresponding frequency. In the case of figure \ref{fig:MTF}, the 10\,\% value of the MTF is 0.54\,lp/mm, which corresponds to two distinguishable slits at 1.85\,mm from each other. For the 3\,\% limit of the MTF, the position resolution improves to 1.67\,mm.

\begin{figure}[tbp]
\centering
\includegraphics[width=.3\textwidth]{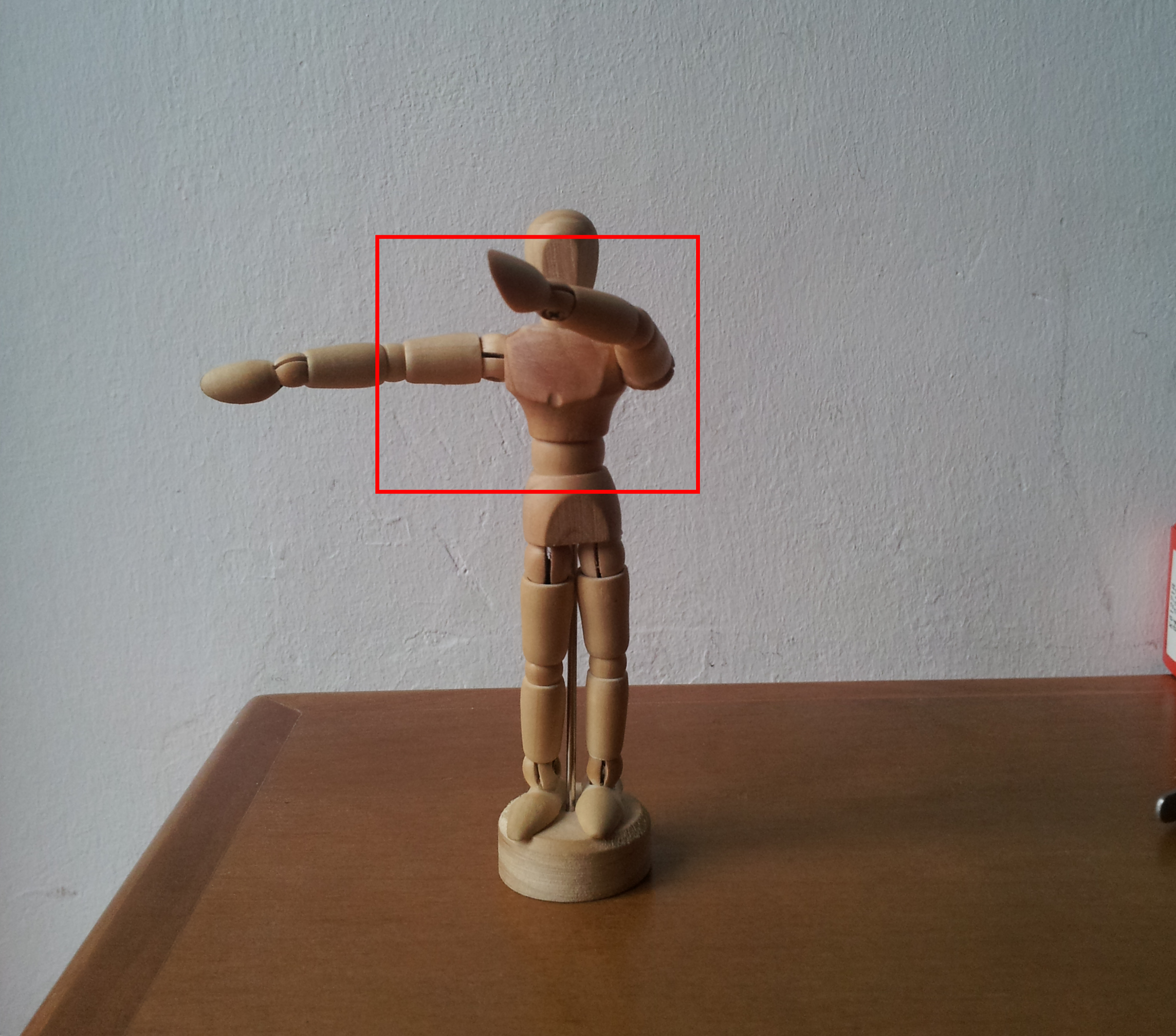}
\includegraphics[width=.5\textwidth]{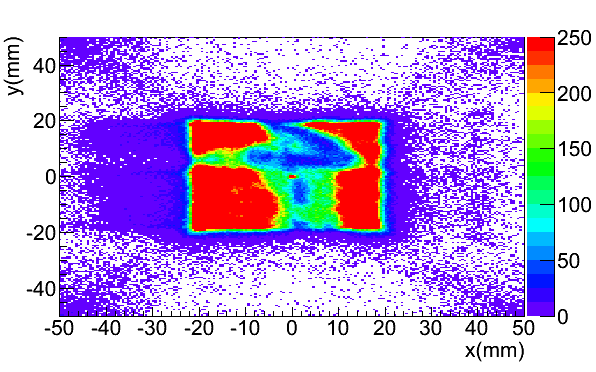}
\caption{The image of a wooden dummy that has a metallic spring as `skeleton'. Both the wooden body (in tones of green) and the metallic spring (in tones of blue - less X-ray trasmission) can be distinguished.}
\label{fig:dum}
\end{figure}

Finally, to test the dynamic range of the imaging system, an articulated wooden dummy, with a metallic spring inside as `skeleton' was imaged. As seen in figure \ref{fig:dum}, both the metallic spring and the dummy's wooden body can be distinguished. This contrast can be changed by selecting different energy ranges. This is one major advantage of this kind of imaging approach, since for each event has its position recorded as well as its energy.

\begin{figure}[tbp]
\centering
\includegraphics[width=.6\textwidth]{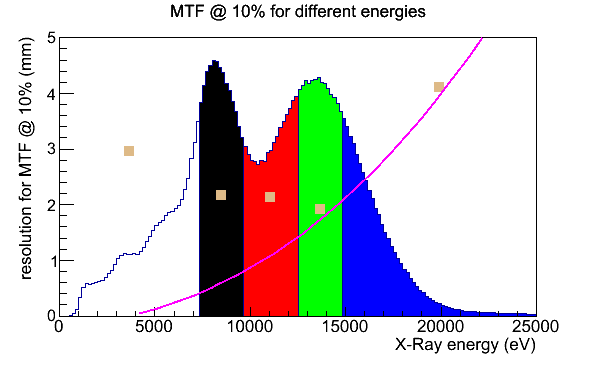}
\caption{The position resolution as a function of the energy (brown squares). The energy spectrum is drawn for reference (the copper peak is seen, strongly attenuated by the 1.5 stainless steel sheet). The magenta line is the photo-electron range in argon as a function of the incoming X-ray energy, from calculated data \cite{Smi05}.}
\label{fig:ResVsE}
\end{figure}

The position resolution improves with increasing photon energy up to 14 keV. This is consistent with the increase of the SNR. In fact, if the value of the 10\,\% limit of the MTF is plotted as a function of the X-ray energy range used to build it, it is evident that the position resolution keeps improving, as shown in figure \ref{fig:ResVsE}. However, the resolution is also influenced by the range of the photo-electrons in the gas mixture (magenta line in the figure). The energy of the K-edge of argon is \keV[3.2], meaning that a \keV[15] X-ray gives origin to a \keV[11.8] photo-electron, which has a range just below 2\,mm \cite{Smi05}. This range has a direct influence on the position resolution and keeps increasing with the energy of the incoming X-rays. The figure shows that for X-ray energies above around \keV[12], the position resolution limitation is related to the gas mixture that was used.

To improve the position resolution, one must increase the SNR to use the lower energies of the spectrum ($< \keV[8]$). Although the noise cannot be reduced \emph{ad infinitum} due to the use of resistive charge division, there is still room for improvements, by better matching of the input capacitances of the charge sensitive preamplifiers. 

\section{Conclusion}\label{sec:conclusion}

An imaging system used in a double GEM configuration with resistive readout has been successfully built, leading to position resolutions of 1.7\,mm for an active area of $5\times 5\,\mathrm{cm}^2$. The charge readout by resistive charge division allowed using more cost effective electronics, with only two charge readout channels for each dimension. The position resolution achieved is in agreement with the expected range of the photo-electrons in argon mixtures for \keV[10] X-rays. Developments for reduction of the signal-to-noise ratio are ongoing, which will allow to use the lower energies of the spectrum, with a subsequent improvement of the position resolution.

The fact that a higher frequency of discharge is not damaging the detector allows us to operate at gains close to the sparking limit is also helping to achieve better results. While testing the detector, some acquisitions were done with the spark rate as high as 0.2\,Hz, with a fast recovery time ($\sim1\,$s), during several minutes. Under such circumstances, we observed no gain instabilities. A systematic study on the spark behavior of these GEMs will be described in a separate work. Nevertheless, this work has already demonstrated their robustness, being able to withstand many discharges without damage. Studies using a $10\times 10\,\mathrm{cm}^2$ readout, exploiting the full area of the GEMs are also foreseen.

\acknowledgments

H. Natal da Luz is supported by FCT grant SFRH/BPD/66737/2009.

\noindent Work carried out within the R\&D project CERN/FP/123614/2011. This project was developed under the scope of a QREN initiative, UE/FEDER financing, through the COMPETE program - Programa Operacional Factores de Competitividade.

\end{document}